\documentclass[12pt,english,floatfix,nofootinbib,superscriptaddress,aps,prd,preprint]{revtex4}
\usepackage[utf8]{inputenc}
\usepackage{float}
\usepackage{array}
\usepackage{bbold}
\usepackage{lipsum}
\usepackage{dsfont}
\usepackage{graphicx}
\usepackage{amsmath,amsthm,amsfonts,amssymb}
\usepackage{graphicx}
\usepackage[english]{babel} 
\usepackage{color}
\usepackage{tensor}
\usepackage{esint}
\usepackage[dvips]{epsfig}
\usepackage[dvips]{graphicx}
\usepackage{float}
\usepackage{units}
\usepackage{textcomp}
\usepackage{mathrsfs}
\usepackage{amsmath}
\usepackage[makeroom]{cancel}
\usepackage{amssymb}
\usepackage{amsbsy}
\usepackage{amsfonts}
\usepackage{amssymb,mathrsfs,xcolor}
\usepackage{esint}
\usepackage{braket}
\usepackage{array}
\usepackage{graphicx}

\usepackage{wasysym}
\usepackage{multirow}
\usepackage{wrapfig}
\usepackage{subfig}

\usepackage{stmaryrd}
\usepackage{upgreek}

\makeatletter

\makeatletter\usepackage{babel}

\usepackage{hyperref}
\hypersetup{
    colorlinks,
    citecolor=blue,
    filecolor=green,
    linkcolor=purple,
    urlcolor=red,
}

\usepackage{slashed}

\newcommand{\ie}{\begin{equation}}
\newcommand{\fe}{\end{equation}}
\newcommand{\se}{\begin{eqnarray}}
\newcommand{\ff}{\end{eqnarray}}

\begin{document}

\title{Comment on ``Quantum tunneling from Schwarzschild black hole in non-commutative gauge theory of gravity''}

%%%%%%%%%%%%%%%%%%%%%%%%%%%%%%%%%%%%%%%%%%%%%%%%%%%%%%%%%%%%%%%%%%%%%%
\author{A. A. Ara\'{u}jo Filho}
\email{dilto@fisica.ufc.br}

\affiliation{Departamento de Física, Universidade Federal da Paraíba, Caixa Postal 5008, 58051-970, João Pessoa, Paraíba,  Brazil.}

%%%%%%%%%%%%%%%%%%%%%%%%%%%%%%%%%%%%%%%%%%%%%%%%%%%%%%%%%%%%%%%%%%%%%%

\author{Iarley P. Lobo}
\email{lobofisica@gmail.com}

\affiliation{Department of Chemistry and Physics, Federal University of Para\'iba, Rodovia BR 079 - km 12, 58397-000 Areia-PB,  Brazil}

\affiliation{Physics Department, Federal University of Campina Grande, Caixa Postal 10071, 58429-900, Campina Grande-PB, Brazil}

%%%%%%%%%%%%%%%%%%%%%%%%%%%%%%%%%%%%%%%%%%%%%%%%%%%%%%%%%%%%%%%%%%%%%%%%%%%%%%%%%%%%%%%%%%%%%%%%%%%%%%%%%%%%%%%%%%%%%%%%%%%%%%%%%%%%%%%%%%%%%%%%%%%%%%%%%%%%%%%%%%%%%%%%%%%%%%%%%%%%%%%%%%%%%%%%%%%%%%%%%%%%%%%%%%%%%%%%%%%%%%%%%%%%%%%%%%%%%%%%%%%%%%%%%%%%%%%%%%%%%%%%%%%%%%%%%%%%%%%%%%%%%%%%%%%%%%%%%%%%%%%%%%%%

\date{\today}

\begin{abstract}

The particle creation via quantum tunneling was recently calculated for the Schwarzschild non--commutative black hole solution in Ref. [Phys. Lett. B 848 (2024) 138335, e-Print: 2310.02445 [gr-qc]]. Nevertheless, it contains inconsistencies in the calculations that need to be properly corrected. In particular, the event horizon was incorrectly determined in that work, which affected all the subsequent calculations. Moreover, the same issue have been repeated elsewhere by the same authors in Refs. \cite{Touati:2023joy,Touati:2021eem,Touati:2022kuf,Touati:2022zbm}.

\end{abstract}

%\keywords{*****}
\maketitle

%%%%%%%%%%%%%%%%%%%%%%%%%%%%%%%%%%%%%%%%%%%%%%%%%%%%%%%%%%%%%%%%%%%%%%%%%%%%%%%%%%%%%%%%%%%%%%%%%%%%%%%%%%%%%%%%%%%%%%%%%%%%%%%%%%%%%%%%%%%%%%%%%%%%%%%%%%%%%%%%%%%%%%%%%%%%%%%%%%%%%%%%%%%%%%%%%%%%%%%%%%%%%%%%%%%%%%%%%%%%%%%%%%%%%%%%%%%%%%%%%%%%%%%%%%%%%%%%%%%%%%%%%%%%%%%%%%%%%%%%%%%%%%%%%%%%%%%%%%%%%%%%%%%%%%%%%%%%%%%%%%%%%%%%%%%%%%%%%%%%%%%%%%%%%%%%%%%%%%%%%%%%%%%%%%%%%%%%%%%%%%%%%%%%%%%%%%%%%%%%%%%%%%%%%

{\section{The general remarks}}

The paper  ``Quantum tunneling from Schwarzschild black hole in non-commutative gauge theory of gravity'' \cite{Touati:2023cxy} explores aspects of particle creation of a Schwarzschild solution within the context of a non--commutative gauge theory via the quantum tunneling process, following a methodology similar to that used in Refs. \cite{Calmet:2023gbw,12aa2025particle,12araujo2024particle,12araujo2025does}. However, the paper contains inconsistencies in the calculations that require correction. This comment aims to identify and address these issues.

In essence, the authors used a {particular} configuration of the non--commutative matrix ($\Theta=\Theta^{23}=-\Theta^{32}$)
\begin{equation}
	\Theta^{\mu\nu}=\left(\begin{matrix}
		0	& 0 & 0 & 0 \\
		0	& 0 & \Theta & 0 \\
		0	& -\Theta & 0 & 0 \\
		0	& 0 & 0 & 0
	\end{matrix}
	\right), \qquad \mu,\nu=0,1,2,3,
\end{equation}
for implementing the non--commutativity \footnote{Furthermore, it is worth noting that a more general approach to constructing black holes within the framework of non-commutative gauge theory has recently been proposed in the literature \cite{Juric:2025kjl}.}, which leads to the following metric components
\begin{subequations}
\begin{align}
	-g_{tt}=&\left(1-\frac{2 M}{r}\right)+\left\{\frac{M\left(88 M^2 + Mr\left(-77+15\sqrt{1-\frac{2M}{r}}\right)-8r^2\left(-2+\sqrt{1-\frac{2M}{r}}\right)\right)}{16 r^4(-2M+r)}\right\}\Theta^{2}+\mathcal{O}(\Theta^{4})\\
	%%%%%%%%%%%%%%%%%%%%%%%%%
	g_{rr}=&\left(1-\frac{2 M}{r}\right)^{-1}+\left\{\frac{M\left(12M^2+Mr\left(-14+\sqrt{1-\frac{2M}{r}}\right)-r^2\left(5+\sqrt{1-\frac{2M}{r}}\right)\right)}{8r^2(2M-r)^3}\right\}\Theta^{2}+\mathcal{O}(\Theta^{4})\label{2b}\\
	%%%%%%%%%%%%%%%%%%%%%%%%
	g_{\theta\theta}=&r^{2}+\left\{\frac{M\left(M\left(10-6\sqrt{1-\frac{2M}{r}}\right)-\frac{8M^2}{r}+r\left(-3+5\sqrt{1-\frac{2M}{r}}\right)\right)}{16(-2M+r)^2}\right\}\Theta^{2}+\mathcal{O}(\Theta^{4})\\
	%%%%%%%%%%%%%%%%%%%%%%%%
	g_{\phi\phi}=&r^{2}+\left\{\frac{5}{8}-\frac{3}{8}\sqrt{1-\frac{2M}{r}}+\frac{M\left(-17+\frac{5}{\sqrt{1-\frac{2M}{r}}}\right)}{16r}+\frac{M^2\sqrt{1-\frac{2M}{r}}}{(-2M+r)^2}\right\}\Theta^{2}+\mathcal{O}(\Theta^{4}).
\end{align}
\end{subequations}

In this manner, the authors argued that by considering the solution of $1/g_{rr}$, the event horizon turns out to be \cite{Touati:2023cxy}: 
\begin{equation}
\label{wrongevent}
	r_{h}^{NC}=r_{h}\left[ 1+\frac{3}{8}\left( \frac{\Theta }{r_{h}}\right)^{2} \right],
\end{equation}
where $r_{h} = 2M$. Nevertheless, this statement is incorrect. From Eq. \eqref{2b}, the expression for $1/g_{rr}$ {\bf{does not}} admit an analytical solution. However, by performing a series expansion of $1/g_{rr}$ under the assumption of a small $\Theta$, we obtain:
\ie
\frac{1}{g_{rr}} \approx \left(1-\frac{2 M}{r}\right) +\frac{\Theta ^2 M \left(12 M^2-r^2 \sqrt{\frac{r-2 M}{r}}-14 M r+M r \sqrt{\frac{r-2 M}{r}}-5 r^2\right)}{8 r^4 (2 M-r)},
\fe
which gives rise to ten possible solutions $r_i$ ($i=1,...,10$) that are complex.

For example, by considering the configuration $M = 1$ and $\Theta = 0.01$, there is no real positive root among the solutions, as can be seen as follows:  
$r_{1} = -0.0167168 - 0.0287987i$,  
$r_{2} = -0.0167168 + 0.0287987i$,  
$r_{3} = -0.0167139 - 0.0292817i$,  
$r_{4} = -0.0167139 + 0.0292817i$,  
$r_{5} = 0.0333995 - 0.000238653i$,  
$r_{6} = 0.0333995 + 0.000238653i$,  
$r_{7} = 2.00002 - 0.00750527i$,  
$r_{8} = 2.00002 + 0.00750527i$,  
$r_{9} = 2.00003 - 0.00750488i$,  
$r_{10} = 2.00003 + 0.00750488i$. The mere presence of these complex solutions already disproves the horizon radius \eqref{wrongevent}.

Moreover, if we take $r_{10}$—the outermost root—and perform a proper expansion assuming small $\Theta$, we get a shorter expression for it
\ie
\label{truehorizon}
r_{10} = 2 M+ \frac{3 i \Theta }{4}+\frac{\Theta ^{3/2}}{32 \sqrt{6} \sqrt{i M}}+\frac{5 \Theta ^2}{32 M},
\fe
which clearly deviates from the result presented by the authors in Ref. \cite{Touati:2023cxy}, specifically in Eq. (\ref{wrongevent}). A direct comparison reveals that the authors neglected the second and third terms of the expansion, which are responsible for the imaginary contributions. In addition to this omission, the final term slightly differs, displaying a factor of “5” instead of the “3” found in Eq. (\ref{wrongevent}).

%Moreover, the only way to reproduce the expression given in Eq. (\ref{wrongevent}) is by adopting a different metric—such as the one discussed in Ref. \cite{chaichian2008corrections}—and solving $g_{tt} = 0$.

Therefore, since the analysis of particle creation via quantum tunneling critically depends on the event horizon, {\bf{all calculations}} presented by the authors {\bf{must be revised}} using the correct expression for the event horizon. Unfortunately, even with the implementation of the corrections highlighted above, it remains unclear whether the analysis of particle creation via quantum tunneling would be physically meaningful, given the presence of imaginary parts for all solutions extracted from the expansion of $1/g_{rr}$ ($r_{1}$ through $r_{10}$). This feature might suggest the existence of a regular black hole lacking a physical event horizon.

Furthermore, {\bf{the same problem}} have been repeated elsewhere by the same authors in Refs. \cite{Touati:2023joy,Touati:2021eem,Touati:2022kuf,Touati:2022zbm}.

%%%%%%%%%%%%%%%%%%%%%%%%%%%%%%%%%%%%%%%%%%%%%%%%%%%%%%%%%%%%%%%%%%%%%%%%%%%%%%%%%%%%%%%%%%%%%%%%%%%%%%%%%%%%%%%%%%%%%%%%%%%%%%%%%%%%%%%%%%%%%%%%%%%%%%%%%%%%%%%%%%%%%%%%%%%%%%%%%%%%%%%%%%%%%%%%%%%%%%%%%%%%%%%%%%%%%%%%%%%%%%%%%%%%%%%%%%%%%%%%%%%%%%%%%%%%%%%%%%%%%%%%

{\section{The corrected solution via $\partial_{r} \wedge \partial_{\theta}$ twist}}

{

We revisit the analysis previously presented in Ref. [Phys. Lett. B 848 (2024) 138335, e-Print: 2310.02445 [gr-qc]] by introducing corrections derived from recent developments reported in \cite{Juric:2025kjl}. The approach adopted here is grounded in a refined framework that incorporates non--commutative deformations at the level of spacetime geometry.

The construction begins by specifying the line element in a $(3+1)$--dimensional non--commutative Schwarzschild background, defined by the metric $\mathrm{d}s^{2} = g_{\mu\nu}(x,\Theta)\mathrm{d}x^{\mu}\mathrm{d}x^{\nu}$, where the coordinates are given by $x^{\mu} = (t, r, \theta, \varphi)$. Within this geometric setup, the determination of the deformed tetrad fields $\hat{{e}}^{a}_{\mu}(x,\Theta)$ becomes essential. These fields are constructed through the contraction of the non--commutative gauge group $\mathrm{SO}(4,1)$ to the Poincaré group $\mathrm{ISO}(3,1)$, using the Seiberg--Witten map as a guiding prescription \cite{3,4,5}.

Moreover, we can define the non--commutative spacetime as follows
\ie
\label{NonCommSTcond1}
\left[x^{\mu},x^{\nu}\right]=i\Theta^{\mu\nu}.
\fe

In non--commutative gravity frameworks, the deformation is introduced via a real antisymmetric tensor $\Theta^{\mu\nu}$, satisfying $\Theta^{\mu\nu} = -\Theta^{\nu\mu}$. This antisymmetric structure serves as an essential tool for constructing modified geometrical quantities. Specifically, both the deformed tetrads $\hat{{e}}^{a}_{\mu}(x,\Theta)$ and the associated spin connection $\hat{\Tilde{\omega}}^{ab}_{\mu}(x,\Theta)$ are developed through a systematic perturbative expansion in powers of $\Theta$. This procedure, established in foundational works \cite{1,2,3,4}, allows one to obtain the corrections induced by non--commutativity in a power series representation
\ie
\begin{split}
		&\hat{{e}}^{a}_{\mu}(x,\Theta) = {e}^{a}_{\mu}(x) - i \Theta^{\nu\rho}{e}^{a}_{\mu\nu\rho}(x) + \Theta^{\nu\rho}\Theta^{\lambda\tau}{e}^{a}_{\mu\nu\rho\lambda\tau}(x)\dots,\\	
		&\hat{\Tilde{\omega}}^{ab}_{\mu}(x,\Theta) = \Tilde{\omega}^{ab}_{\mu}(x)-i \Theta^{\nu\rho} \Tilde{\omega}^{ab}_{\mu\nu\rho}(x) + \Theta^{\nu\rho}\Theta^{\lambda\tau}\Tilde{\omega}^{ab}_{\mu\nu\rho\lambda\tau}(x)\dots  \,\,. \label{omeganoncom}
	\end{split}
\fe 
{where, as we will show below, ${e}^{a}_{\mu\nu\rho}(x)$, ${e}^{a}_{\mu\nu\rho\lambda\tau}(x)$, $\Tilde{\omega}^{ab}_{\mu\nu\rho}(x)$ and $\Tilde{\omega}^{ab}_{\mu\nu\rho\lambda\tau}(x)$ are functions of the undeformed tetrads ${e}^{a}_{\mu}$, the spin connection $\Tilde{\omega}^{ab}_{\mu}$, the curvature tensor $R^{ab}_{\mu\nu}$, and the Minkowski metric $\eta_{\mu\nu}$, as explained in \eqref{noncommcorr-tetrad}, \eqref{noncommcorr-omega}, \eqref{correction-e1}, \eqref{3.12}.}

The expansion of the gauge connection $\hat{\Tilde{\omega}}^{ab}_{\mu}(x,\Theta)$ up to second order in the non--commutativity parameter $\Theta$, as outlined in Eq. \eqref{omeganoncom}, forms the basis for deriving the corresponding deformed tetrads $\hat{{e}}^{a}_{\mu}(x,\Theta)$. These tetrad fields incorporate the geometric effects introduced by non--commutative deformations and reflect the modified gauge structure that underlies the construction of gravity in such settings \cite{Juric:2025kjl} 
\ie
\tilde{\omega}^{ab}_{\mu\nu\rho} (x) = \frac{1}{4} \left\{\Tilde{\omega}_{\nu},\partial_{\rho}\Tilde{\omega}_{\mu}+R_{\rho\mu}\right\}^{ab},\label{noncommcorr-tetrad}
\fe
\begin{equation}
	\begin{split}
		\Tilde{\omega}_{\mu\nu\rho\lambda\tau}^{ab} = &\frac{1}{16}\biggl[-\left\{\left\{\Tilde{\omega}_{\lambda},\left(\partial_\tau \Tilde{\omega}_{\nu} + R_{\tau\nu}\right)\right\},\left(\partial_\rho  \Tilde{\omega}_{\mu} + R_{\rho\mu}\right)\right\}^{ab} \\
  & - \left\{\Tilde{\omega}_{\nu}, \partial_\rho \left\{\Tilde{\omega}_{\lambda},\left(\partial_{\tau}\Tilde{\omega}_{\mu} + R_{\tau\mu}\right)\right\} \right\}^{ab}  +2 \left[\partial_{\lambda} \Tilde{\omega}_{\nu}, \partial_{\tau} \left(\partial_{\rho} \Tilde{\omega}_{\mu} + R_{\rho\mu}\right)\right]^{ab}  \\
  & + \left\{\Tilde{\omega}_{\nu},2\left\{R_{\rho\lambda},R_{\mu\tau}\right\}\right\}^{ab} - \left\{\Tilde{\omega}_{\nu},\left\{\Tilde{\omega}_{\lambda}, D_\tau R_{\rho\mu} + \partial_\tau R_{\rho \mu}\right\}\right\}^{ab}   \biggr].
	\end{split}
\label{noncommcorr-omega}
\end{equation}

Under the Seiberg--Witten map, Eqs. \eqref{noncommcorr-tetrad} and \eqref{noncommcorr-omega} must obey the gauge consistency conditions {(which defines the brackets used above)}
\ie\label{seiberg-witten}
\left[\alpha,\beta\right]^{ab} =  \alpha^{ac}\beta^{b}_{c}-\beta^{ac}\alpha^{b}_{c},\qquad
\left\{\alpha,\beta\right\}^{ab} = \alpha^{ac}\beta^{b}_{c}+\beta^{ac}\alpha^{b}_{c},
\fe
and
\ie
	D_{\mu}R^{ab}_{\rho\sigma} = \partial_{\mu}R^{ab}_{\rho\sigma} +\left(\Tilde{\omega}^{ac}_{\mu}R^{db}_{\rho\sigma}+\Tilde{\omega}^{bc}_{\mu}R^{da}_{\rho\sigma}\right)\eta_{cd},
\fe
{where the curvature tensor is
\begin{eqnarray}
    R^{ab}_{\mu\nu}=\partial_{\mu}\tilde{\omega}^{ab}_{\nu}-\partial_{\nu}\tilde{\omega}^{ab}_{\mu}+\eta_{cd}\left(\tilde{\omega}^{ac}_{\mu}\tilde{\omega}^{db}_{\nu}-\tilde{\omega}^{ac}_{\nu}\tilde{\omega}^{db}_{\mu}\right).
\end{eqnarray}}

The gauge connection $\hat{\Tilde{\omega}}^{ab}_{\mu}(x,\Theta)$ is subject to specific constraints, which must be respected to preserve the underlying conditions
\ie
\label{CondDefOmega}
\hat{\Tilde{\omega}}^{ab\star}_{\mu}(x,\Theta) = -\hat{\Tilde{\omega}}^{ab}_{\mu}(x,\Theta), \quad 
\hat{\Tilde{\omega}}^{ab}_{\mu}(x,\Theta) ^{r} \equiv \hat{\Tilde{\omega}}^{ab}_{\mu}(x,-\Theta) = -\hat{\Tilde{\omega}}^{ba}_{\mu}(x,\Theta).
\fe

Note that the symbol ${}^\star$ denotes complex conjugation. In addition, the non--commutative corrections resulting from the constraints in Eq. \eqref{CondDefOmega} take the form:
\ie
\Tilde{\omega}^{ab}_{\mu} (x) = - \Tilde{\omega}^{ba}_{\mu} (x), \quad \Tilde{\omega}^{ab}_{\mu\nu\rho} (x) = \Tilde{\omega}^{ba}_{\mu\nu\rho} (x), \quad \Tilde{\omega}^{ab}_{\mu\nu\rho\lambda\tau} (x) = -\Tilde{\omega}^{ba}_{\mu\nu\rho\lambda\tau} (x).
\fe

{The expressions above result from applying Eqs. \eqref{noncommcorr-tetrad} and \eqref{noncommcorr-omega}, imposing the torsionless condition $T^{a}_{\mu\nu} = 0$ and taking the limit ${K} \to 0$ (where $K$ is the contraction parameter that reduces the gauge group to ISO(3,1) in this limit).} In this regime, one obtains:
\ie
\label{ComConjDefTetrads}
\hat{{e}}^{a\star}_{\mu}(x,\Theta) = {e}^{a}_{\mu}(x)+i \Theta^{\nu\rho}{e}^{a}_{\mu\nu\rho}(x)+\Theta^{\nu\rho}\Theta^{\lambda\tau}{e}^{a}_{\mu\nu\rho\lambda\tau}(x)\dots,
\fe
in which
\ie
\begin{split}\label{correction-e1}
{e}^{a}_{\mu\nu\rho} &= \frac14	\left[\Tilde{\omega}^{ac}_{\nu}\partial_{\rho} {e}^{d}_{\mu} + \left(\partial_{\rho}\Tilde{\omega}^{ac}_{\mu} + R^{ac}_{\rho\mu}\right){e}^{d}_{\nu}\right]\eta_{cd},
\end{split}
\fe
with \cite{Juric:2025kjl}
\begin{align}
&{e}_{\mu \nu \rho \lambda \tau }^{a} = \nonumber\\
&= \frac{1}{16}\Bigl[
   2\,\Bigl\{R_{\tau \nu},\,R_{\mu \rho}\Bigr\}^{ab}\,{e}_{\lambda}^{c}
   \;-\;\Tilde{\omega}_{\lambda}^{a\,b}\,\Bigl(D_{\rho}\,R_{\tau \mu}^{c\,d}
     \;+\;\partial_{\rho}\,R_{\tau \mu}^{c\,d}\Bigr)\,{e}_{\nu}^{m}\,\eta_{d\,m}
   -\,\Bigl\{\Tilde{\omega}_{\nu},\,\bigl(D_{\rho}\,R_{\tau \mu}
     + \partial_{\rho}\,R_{\tau \mu}\bigr)\Bigr\}^{ab}\,{e}_{\lambda}^{c}
     \nonumber \\%[1pt]
&\quad
   \;-\;\partial_{\tau}\,\Bigl\{\Tilde{\omega}_{\nu},\,\bigl(\partial_{\rho}\,\Tilde{\omega}_{\mu}
     + R_{\rho \mu}\bigr)\Bigr\}^{a\,b}\,{e}_{\lambda}^{c}
   -\,\Tilde{\omega}_{\lambda}^{a\,b}\,\partial_{\tau}\Bigl(
       \Tilde{\omega}_{\nu}^{c\,d}\,\partial_{\rho}\,{e}_{\mu}^{m}
       + \bigl(\partial_{\rho}\,\Tilde{\omega}_{\mu}^{c\,d}
         + R_{\rho \mu}^{c\,d}\bigr)\,{e}_{\nu}^{m}
     \Bigr)\,\eta_{d\,m}
     \nonumber \\%[1pt]
&\quad
   \;+\;2\,\partial_{\nu}\,\Tilde{\omega}_{\lambda}^{a\,b}\,
        \partial_{\rho}\partial_{\tau}\,{e}_{\mu}^{c}
   -\,2\,\partial_{\rho}\Bigl(\partial_{\tau}\,\Tilde{\omega}_{\mu}^{a\,b}
     + R_{\tau \mu}^{a\,b}\Bigr)\,\partial_{\nu}\,{e}_{\lambda}^{c}
   \;-\;\Bigl\{\Tilde{\omega}_{\nu},\,\bigl(\partial_{\rho}\,\Tilde{\omega}_{\lambda}
     + R_{\rho \lambda}\bigr)\Bigr\}^{a\,b}\,\partial_{\tau}\,{e}_{\mu}^{c}
\nonumber \\%[1pt]
&\quad
   \;-\,\Bigl(\partial_{\tau}\,\Tilde{\omega}_{\mu}^{a\,b}
     + R_{\tau \mu}^{a\,b}\Bigr)\,\Bigl(
       \Tilde{\omega}_{\nu}^{c\,d}\,\partial_{\rho}\,{e}_{\lambda}^{m}
       + \bigl(\partial_{\rho}\,\Tilde{\omega}_{\lambda}^{c\,d}
         + R_{\rho \lambda}^{c\,d}\bigr)\,{e}_{\nu}^{m}\,\eta_{d\,m}
     \Bigr)
\Bigr]\;\eta_{b\,c}
\nonumber \\%[1pt]
&\quad
\; - \frac{1}{16}\,\Tilde{\omega}_{\lambda}^{a\,c}\,\Tilde{\omega}_{\nu}^{d\,b}\,{e}_{\rho}^{f}\,
   R_{\tau \mu}^{g\,m}\,\eta_{c\,d}\,\eta_{f\,g}\,\eta_{b\,m}\,.
\label{3.12}
\end{align}

This final contribution was not included in the original analysis of Ref. \cite{chaichian2008corrections}, nor in later works that extended its approach. Consequently, the corrected metric tensor assumes the following form \cite{Heidari:2025iiv}:
\ie
\label{DefMetTensor}
\hat{g}_{\mu\nu}\left(x,\Theta\right) = \frac{1}{2} \eta_{ab}\Bigg[\hat{{e}}^{a}_{\mu}(x,\Theta)\ast\hat{{e}}^{b\star}_{\nu}(x,\Theta)+\hat{{e}}^{b}_{\mu}(x,\Theta)\ast\hat{{e}}^{a\star}_{\nu}(x,\Theta)\Bigg].
\fe

Here, the symbol $\ast$ refers to the standard star product \cite{Juric:2025kjl}. For the calculations that follow, we adopt natural units by setting $\hbar = c = G = 1$. Within this convention, and using the same twist employed in earlier analyses—specifically the Moyal product defined via $\partial_{r} \wedge \partial_{\theta}$ \cite{Juric:2025kjl}—the resulting expression is given by:
\begin{equation}
    \begin{split}
        \hat{g}_{tt}\left( x,\Theta \right) &= -\left(1 - \frac{2 M}{r} \right)+ \frac{M (11 M - 4 r)}{2 r^4}  \Theta^2 + \mathcal{O}(\Theta^3), \\
        \hat{g}_{rr}\left( x,\Theta \right) &= \left(1 -  \frac{2 M}{r}\right)^{-1} + \frac{M (3 M - 2 r) }{2 r^2 (r-2 M)^2} \Theta^2 + \mathcal{O}(\Theta^3), \\
        \hat{g}_{\theta\theta}\left( x,\Theta \right) &= r^2 + \left(\frac{1}{16} -  \frac{2 M}{r} + \frac{M}{8 (r-2 M )}\right) \Theta^2 + \mathcal{O}(\Theta^3), \\
        \hat{g}_{\varphi\varphi}\left( x,\Theta \right) &= r^2 \sin ^2\theta + \left(\frac{5 \cos ^2\theta }{16}+\frac{\sin ^2\theta  \left(2 M^2-4 M r+r^2\right)}{4 r (r-2 M)}\right)\Theta ^2 +\mathcal{O}\left(\Theta ^3\right).    \end{split}
\end{equation}

}

%%%%%%%%%%%%%%%%%%%%%%%%%%%%%%%%%%%%%%%%%%%%%%%%%%%%%%%%%%%%%%%%%%%%%%%%%%%%%%%%%%%%%%%%%%%%%%%%%%%%%%%%%%%%%%%%%%%%%%%%%%%%%%%%%%%%%%%%%%%%%%%%%%%%%%%%%%%%%%%%%%%%%%%%%%%%%%%%%%%%%%%%%%%%%%%%%%%%%%%%%%%%%%%%%%%%%%%%%%%%%%%%%%%%%%%%%%%%%%%%%%%%%%%%%%%%%%%%%%%%%%%%%

{\section{The quantum tunneling process}}

\begin{figure}
    \centering
    \includegraphics[scale=0.6]{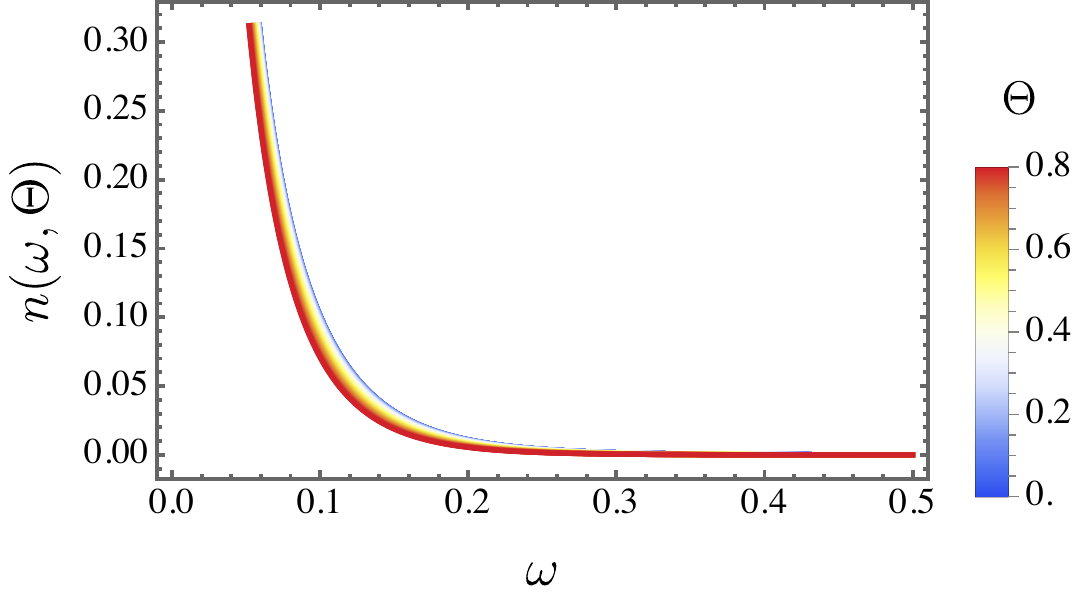}
    \caption{The particle creation density $n(\omega,\Theta)$ is plotted as a function of the frequency $\omega$, with curves corresponding to distinct values of the non--commutative parameter $\Theta$.}
    \label{bosonparticles}
\end{figure}

{

To account for energy conservation in the radiation spectrum, we follow the method developed in \cite{011, vanzo2011tunnelling, parikh2004energy, Calmet:2023gbw}. By rewriting the metric in the Painlevé--Gullstrand form, it takes the structure
$\mathrm{d}s^2 = - f(r,\Theta)\,\mathrm{d}t^2 + 2 h(r,\Theta) \,\mathrm{d}t \mathrm{d}r + \mathrm{d}r^2 + z(r,\theta,\Theta)\,\mathrm{d}\theta^2 + y(r,\theta,\Theta)\,\mathrm{d}\varphi^2,$
with $h(r) = \sqrt{f(r)\big(g(r)^{-1} - 1\big)}$ as derived in \cite{Calmet:2023gbw}. The tunneling probability is then determined by the imaginary part of the action \cite{parikh2004energy, vanzo2011tunnelling, Calmet:2023gbw}. The classical action for a massless particle is given by
$\mathcal{S} = \int p_\mu \, \mathrm{d}x^\mu.$

When computing $\text{Im}\,\mathcal{S}$, only the radial component contributes to the imaginary part, since $p_t \mathrm{d}t = -\omega \mathrm{d}t$ is purely real and does not affect it. Thus, the relevant contribution comes exclusively from:
\ie
\text{Im}\,\mathcal{S}=\text{Im}\,\int_{r_i}^{r_f} \,p_r\,\mathrm{d}r=\text{Im}\,\int_{r_i}^{r_f}\int_{0}^{p_r} \,\mathrm{d}p_r'\,\mathrm{d}r.
\fe

Considering the Hamiltonian $H = M - \omega'$, where $\omega'$ represents the instantaneous energy of the radiated particle, Hamilton’s relation gives $\mathrm{d}H = -\mathrm{d}\omega'$. With the energy varying from $0$ to $\omega$, the total emitted energy, the resulting contribution to the imaginary part of the action is expressed as:
\ie
\begin{split}
\text{Im}\, \mathcal{S} & = \text{Im}\,\int_{r_i}^{r_f}\int_{M}^{M-\omega} \,\frac{\mathrm{d}H}{\mathrm{d}r/\mathrm{d}t}\,\mathrm{d} r  =\text{Im}\,\int_{r_i}^{r_f}\,\mathrm{d}r\int_{0}^{\omega} \,-\frac{\mathrm{d}\omega'}{\mathrm{d}r/\mathrm{d}t}\,.
\end{split}
\fe
After changing the order of integration and introducing the appropriate substitution, the expression becomes:
\ie
\begin{split}
\frac{\mathrm{d}r}{\mathrm{d}t} =&  -h(r)+\sqrt{f(r)+h(r)^2} \\
& = \frac{1}{2} \left(\sqrt{\frac{\left(2 r^3 (r-2 M)+\Theta ^2 M (11 M-4 r)\right) \left(2 r^3 (r-2 M)+\Theta ^2 M (3 M-2 r)\right)}{r^6 (r-2 M)^2}} \right.\\
& \left.  -\sqrt{-\frac{M \left(2 r^3 (r-2 M)+\Theta ^2 M (11 M-4 r)\right) \left(-3 \Theta ^2 M+8 M r^2-4 r^3+2 \Theta ^2 r\right)}{r^6 (r-2 M)^2}}\right)\\
& \approx  \, \left(1-\sqrt{2} \sqrt{\frac{M}{r}}\right) + \frac{1}{2} \Theta ^2 \left(\frac{7 M^2-3 M r}{r^3 (r-2 M)}+\frac{\sqrt{M} \left(-22 M^2+5 M r+2 r^2\right)}{2 \sqrt{2} r^{7/2} (r-2 M)}\right),
\end{split}
\fe
up to the second order in $\Theta$. Now, we implement the replacement $M \to (M - \omega')$ in the metric, leading to a modified expression for the function, given by:
\ie
\begin{split}
\label{ims}
&\text{Im}\, \mathcal{S} =\text{Im}\,\int_{0}^{\omega} -\mathrm{d}\omega'\\
& \times \int_{r_i}^{r_f}\,\frac{\mathrm{d}r}{\left(1-\sqrt{2} \sqrt{\frac{M-\omega'}{r}}\right)+\frac{1}{2} \Theta ^2 \left[\frac{7 (M-\omega')^2-3 r (M-\omega')}{r^3 (r-2 (M-\omega'))}+\frac{\sqrt{M-\omega'} \left(5 r (M-\omega')-22 (M-\omega')^2+2 r^2\right)}{2 \sqrt{2} r^{7/2} (r-2 (M-\omega'))}\right]}.
\end{split}
\fe

As an example, let us define $\Delta(r,\omega') \equiv 2 (M - \omega')$. Notice that the substitution of mass, i.e., $M \to (M - \omega')$, shifts the location of the pole to the modified horizon at $r = 2(M - \omega')$. Prior to evaluating the integral, we expand the integrand up to second order in the parameter $\Theta$, yielding:
\ie
\begin{split}
\label{imspertubated}
&\text{Im}\, \mathcal{S} \approx \, \text{Im}\,\int_{0}^{\omega} -\mathrm{d}\omega'\\
& \times \int_{r_i}^{r_f}\, 
\left\{ \frac{1}{1-\sqrt{2} \sqrt{\frac{M-\omega'}{r}}}-\frac{\Theta ^2 \left[\frac{7 (M-\omega')^2-3 r (M-\omega')}{r^3 (r-2 (M-\omega'))}+\frac{\sqrt{M-\omega'} \left(5 r (M-\omega')-22 (M-\omega')^2+2 r^2\right)}{2 \sqrt{2} r^{7/2} (r-2 (M-\omega'))}\right]}{2 \left(1-\sqrt{2} \sqrt{\frac{M-\omega'}{r}}\right)^2} \right\} \mathrm{d}r.
\end{split}
\fe

Carrying out the contour integration around this pole in the counterclockwise direction gives:
\begin{eqnarray}
    \text{Im}\, \mathcal{S}  = 2 \pi  \omega  (2 M-\omega )+ \frac{25}{32} \pi  \Theta ^2 \Big[\ln (M)-\ln (M-\omega )\Big] .
\end{eqnarray}

In this manner, the emission rate for a \textit{Hawking} particle, including the effects of non--commutative corrections, takes the form:
\ie
\Gamma (\omega,\Theta) \sim e^{-2 \, \text{Im}\, S} = e^{- 4 \pi  \omega  (2 M-\omega ) - \frac{25}{16} \pi  \Theta ^2 \Big[\ln (M)-\ln (M-\omega )\Big] } .
\fe
In addition, the particle number density can be expressed through the tunneling rate as follows:
\ie
n(\omega,\Theta) = \frac{\Gamma(\omega,\Theta)}{1 - \Gamma(\omega,\Theta)} = \frac{1}{\exp \left\{\frac{25}{16} \pi  \Theta ^2 [\ln M -\ln (M-\omega )]+4 \pi  \omega  (2 M-\omega )\right\}-1}.
\fe

To illustrate the behavior of $n(\omega,\Theta)$, Fig. \ref{bosonparticles} displays its variation with respect to the non--commutative parameter $\Theta$. It is evident that increasing $\Theta$ suppresses the corresponding particle creation rate. Moreover, we do not revisit the calculation of the Hawking temperature presented in [Phys. Lett. B 848 (2024) 138335, e-Print: 2310.02445 [gr-qc]], since the methodology recently developed in \cite{Juric:2025kjl} reveals that the surface gravity is not well--defined under the specific twist configuration adopted here, namely $\partial_{r} \wedge \partial_{\theta}$.
This indicates that, in this non--commutative scenario, black hole thermodynamics should not be done using the usual geometric definitions.}

%%%%%%%%%%%%%%%%%%%%%%%%%%%%%%%%%%%%%%%%%%%%%%%%%%%%%%%%%%%%%%%%%%%%%%%%%%%%%%%%%%%%%%%%%%%%%%%%%%%%%%%%%%%%%%%%%%%%%%%%%%%%%%%%%%%%%%%%%%%%%%%%%%%%%%%%%%%%%%%%%%%%%%%%%%%%%%%%%%%%%%%%%%%%%%%%%%%%%%%%%%%%%%%%%%%%%%%%%%%%%%%%%%%%%%%%%%%%%%%%%%%%%%%%%%%%%%%%%%%%%%%%%

\section*{Acknowledgments}
\hspace{0.5cm}

A. A. Araújo Filho acknowledges support from the Conselho Nacional de Desenvolvimento Científico e Tecnológico (CNPq) and the Fundação de Apoio à Pesquisa do Estado da Paraíba (FAPESQ) under grant [150891/2023-7]. I. P. L. was partially supported by the National Council for Scientific and Technological Development - CNPq grant 312547/2023-4. I. P. L. would like to acknowledge networking support by the COST Action BridgeQG (CA23130) and by the COST Action RQI (CA23115), supported by COST (European Cooperation in Science and Technology). The authors also express gratitude to N. Heidari for the useful discussions.

%%%%%%%%%%%%%%%%%%%%%%%%%%%%%%%%%%%%%%%%%%%%%%%%%%%%%%%%%%%%%%%%%%%%%%%%%%%%%%%%%%%%%%%%%%
\section{Data Availability Statement}

Data Availability Statement: No Data associated in the manuscript

%%%%%%%%%%%%%%%%%%%%%%%%%%%%%%%%%%%%%%%%%%%%%%%%%%%%%%%%%%%%%%%%%%%%%%%%%%%%%%%%%%%%%%%%%%

\bibliographystyle{ieeetr}
\bibliography{main}

\end{document}